# APPLICATION OF THEOREMS ON NULL-GEODESICS ON THE SOLAR LIMB EFFECT

F.I. MIKHAIL[1], M.I. WANAS[2] and A.B. MORCOS[3,**]

[1] *Late Professor of Mathematics, Ain Shams University, Cairo, Egypt*
[2] *Astronomy Department, Faculty of Science, Cairo University, Giza, Egypt*
E-mail: wanas@mailer.scu.eun.eg
[3] *Department of Physics and Astronomy, University of Leeds, U.K.*
E-mail: phyabm@phys-irc.novell.leeds.ac.uk, or a.morcos@excite.com

**Abstract.** A direct and more general calculation of the limb effect connected with red-shift observations is obtained. Result agrees completely, in its general form, with that obtained from observations of the solar spectra but with different value for the maximum effect.

## 1. Introduction

More consideration has been given recently to the so called 'limb effect' in connection with the red-shift observed for the spectra of the Sun. It has been realized, on an observational basis, that the value of red-shift increases as we move towards the limb of the Sun's disc. Several authors gave possible interpretations for this phenomena. They suggested that the reason may be due to streaming motion in the solar atmosphere, (Adam, 1979), acoustic waves, (Adam, 1976), Compton effect, (Adam, 1976), inter atomic collision, (Adam, 1976), variation of intensity from center to limb, (Peter, 1999), gas flow or waves flow downward from corona to the center, (Mariska, 1992; Hansteen, 1993, radiative transfer; Warren et al., 1998), the variation of magnetic field, (Brynildsen et al., 1995, variation of the electron temperature; Achour et al., 1995; Peter and Judge, 1999, and the scattered light at the extreme limb, Adam, 1976).

It is the purpose of the present work to calculate, in a more general way, the value of the red-shift due to the gravitational field of a celestial body (the Sun say). In doing so, we hope to throw more light on this phenomenon.

** Permanent address: National Research Institute of Astronomy and Geophysics, Department of Astronomy, P.O. Box 471, Maadi, Code No. 11728, Cairo, Egypt.

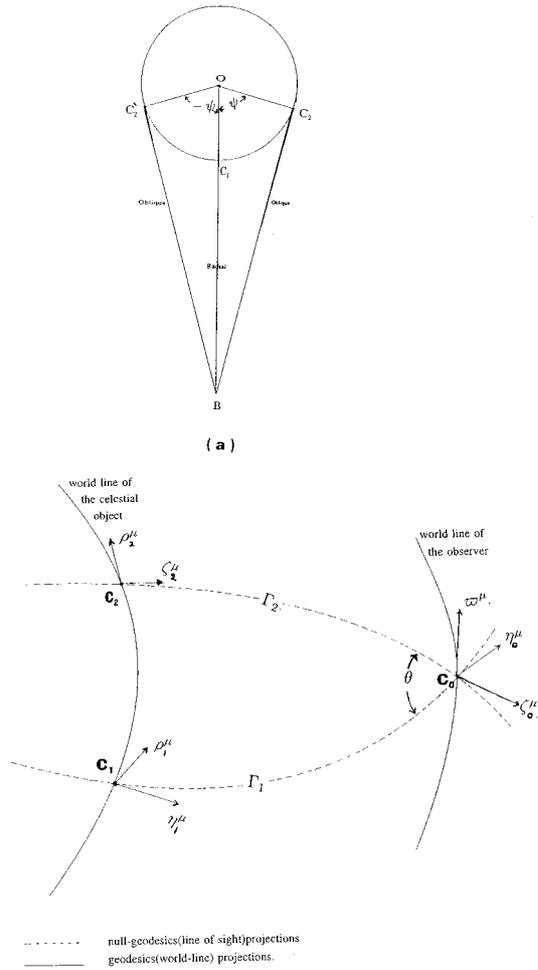

*Figure 1.*

## 2. Theorems on Null-Geodesics and General Formula for Red-Shift

Let us consider two identical atoms at the two points $C_1$, $C_2$ on the equator of a spherical celestial body whose centre is O. The two atoms are accompanied by two observers $A_1$, $A_2$. Thus the two observers $A_1$, $A_2$ will have the same world line of $C_1$, $C_2$. Let the unit vector along their common world line be denoted by $\rho^\mu$ as shown in Figure 1b. Let B be an observer moving in the field of that celestial body (e.g on the surface of the Earth), with the directional unit vector $\varpi^\mu$. At a certain instant, let the observer B be at the point $C_0$ which lies along the radial null-geodesic $\Gamma_1$ passing through $C_1$. Let the null-geodesic joining $C_0$ and $C_2$ be $\Gamma_2$. If $\lambda_1$ is the wavelength of vibration of the first atom at $C_1$ as measured by its accompanying observer $A_1$ and $\underset{0}{\lambda_1}$ is the wavelength, of the same vibration, as measured by the observer B. It has been found by Kermack, McCrea and Whittaker (1933), as a consequence of applying their two theorems on null-geodesics, that

$$\underset{0}{\lambda_1} = \frac{[\rho_\mu \eta^\mu]_{C_1}}{[\varpi_\mu \eta^\mu]_{C_0}} \lambda_1, \tag{1}$$

where $\eta^\mu$ is the transport null-vector along the null-geodesic $\Gamma_1$. The suffixes $C_1$, $C_0$ denote that the expressions within the brackets are to be calculated at the two points $C_1$, $C_0$ respectively.

Similarly, if $\lambda_2$, $\underset{0}{\lambda_2}$ are two wavelengths of vibration of the second atom at $C_2$ as measured by the two observers $A_2$, and B respectively, then we have

$$\underset{0}{\lambda_2} = \frac{[\rho_\mu \zeta^\mu]_{C_2}}{[\varpi_\mu \zeta^\mu]_{C_0}} \lambda_2, \tag{2}$$

where $\zeta^\mu$ is the transport null-vector along the null-geodesic $\Gamma_2$. But, since the two atoms are identical and the celestial body, on its surface they lie, is spherical, the two wavelengths $\lambda_1$, $\lambda_2$ must be identical as the two atoms are on the same great circle, i.e.

$$\lambda_1 = \lambda_2 = \lambda. \tag{3}$$

Thus, the amount of red-shift which is purely due to the change in the line of sight used by the observer will then be given by

$$\Delta Z = \frac{\underset{o}{\lambda_2} - \underset{o}{\lambda_1}}{\lambda}. \tag{4}$$

Using (1),(2), we get

$$\Delta Z = \frac{[\rho_\mu \eta^\mu]_{C_1}}{[\varpi_\mu \eta^\mu]_{C_0}} - \frac{[\rho_\mu \zeta^\mu]_{C_2}}{[\varpi_\mu \zeta^\mu]_{C_0}}. \tag{5}$$

It is then left to calculate the values of the transport vectors $\eta^\mu$, $\zeta^\mu$ along the two null-geodesics $\Gamma_1$, $\Gamma_2$ respectively and the values of the unit vectors $\rho^\mu$, $\varpi^\mu$ at the indicated points. To do so, we are going to consider the solutions of the equation of motion applicable for material particles as well as for photons, i.e. the solutions for the geodesic equation and the null-geodesic equation respectively, in the field of the Sun.

## 3. General Form for The Solution of Equation of Motion

Assuming the exterior field of the celestial body to be of the form of the ordinary Schwarzschild's space-time, i.e.

$$dS^2 = \gamma \, dt^2 - \frac{dr^2}{\gamma} - r^2 \, (d\theta^2 + \sin^2\theta \, d\phi^2), \tag{6}$$

where $\gamma = 1 - \frac{2m}{r}$, $m = \frac{GM}{C^2}$, and M being the mass of the celestial body in the cgs units, the equation of motion of a free test particle can be put (cf. Adler et al., 1975, p. 54) into the form

$$\frac{d^2 x^\mu}{dp^2} + \begin{Bmatrix} \mu \\ \alpha\beta \end{Bmatrix} \frac{dx^\alpha}{dp} \frac{dx^\beta}{dp} = 0 \tag{7}$$

where $p$ is the affine parameter characterizing the trajectory of the particle. The solution of (7) for the metric (6) can be put into the general form,

$$\frac{dr}{dp} = \sqrt{\alpha^2 - \gamma(E + \frac{h^2}{r^2})}, \quad (a)$$

$$\theta = \frac{\pi}{2}, \quad (b)$$

$$\frac{d\phi}{dp} = \frac{h}{r^2}, \quad (c) \tag{8}$$

$$\frac{dt}{dp} = \frac{\alpha}{\gamma}. \quad (d)$$

where $\alpha$, h are two constants of integration, and $E$ is a parameter, specified by

$$E = 0, \text{ for a photon,}$$
$$E = 1, \text{ for a material particle.} \tag{9}$$

It is clear from the Equation (8c), that for a radial path ($\phi = $ constant)

$$\frac{d\phi}{dp} = 0, \text{ and } h = 0, \tag{10}$$

thus from (8)–(10), the components of the transport null-vectors $\eta^\mu$ giving the tangent at any point on the radial null-geodesic $\Gamma_1$, will be

$$\eta^\mu \equiv \left[ \frac{dt}{dp}, \frac{dr}{dp}, \frac{d\theta}{dp}, \frac{d\phi}{dp} \right],$$
$$\equiv \left[ \frac{\alpha}{\gamma}, \alpha, 0, 0 \right]. \tag{11}$$

Thus taking the Sun as our celestial body and the Earth as the point where the observer B is located, the components of the transport null-vector will be:

$$\eta_1^\mu \equiv \left[\frac{\alpha}{\gamma_*}, \alpha, 0, 0\right]. \tag{12}$$

$$\eta_0^\mu \equiv \left[\frac{\alpha}{\gamma_\oplus}, \alpha, 0, 0\right]. \tag{13}$$

at the two points $C_1$ and $C_0$ respectively, where $\gamma_* = 1 - \frac{2m}{a}$, $\gamma_\oplus = 1 - \frac{2m}{b}$, $a$ is the radius of the celestial object (Sun) and $b$ is the distance of the Earth from the the celestial object's (Sun's) center. For the oblique null-geodesic $\Gamma_2$, we still have $E = 0$ while $h \neq 0$. Thus the transport null-vector $\zeta^\mu$ will then have the components:

$$\zeta^\mu \equiv \left[\alpha\, \gamma^{-1}, \sqrt{\alpha^2 - \frac{h^2}{r^2}\gamma}, 0, \frac{h}{r^2}\right], \tag{14}$$

consequently:

$$\zeta_2^\mu \equiv \left[\alpha\, \gamma_*^{-1}, \sqrt{\alpha^2 - \frac{h^2}{a^2}\gamma_*}, 0, \frac{h}{a^2}\right], \tag{15}$$

and

$$\zeta_0^\mu \equiv \left[\alpha\, \gamma_\oplus^{-1}, \sqrt{\alpha^2 - \frac{h^2}{b^2}\gamma_\oplus}, 0, \frac{h}{b^2}\right]. \tag{16}$$

If we assume that the celestial body (Sun) rotates about its center then its angular velocity is given by

$$\frac{d\phi}{dt} \stackrel{\text{def}}{=} \hat{V}_*,$$

i.e.,

$$\frac{d\phi}{dP} = \hat{V}_* \frac{dt}{dP}. \tag{17}$$

By considering a particle on the equator of the celestial object, it moves in a circle of radius equivalent to the radius of the celestial object. So, the unit vector $\rho_\mu$ along the world line of $A_1$ takes the form,

$$\rho_\mu = \left[\frac{\gamma_*}{\sqrt{\gamma_* - \hat{V}_*^2 a^2}}, 0, 0, \frac{-\hat{V}_* a^2}{\sqrt{\gamma_* - \hat{V}_*^2 a^2}}\right]. \tag{18}$$

Also, the unit vector along the world line of the observer at B is given by

$$\varpi_\mu = \left[ \frac{\gamma_\oplus}{\sqrt{\gamma_\oplus - \hat{V}_\oplus^2 b^2}},\ 0,\ 0,\ \frac{-\hat{V}_\oplus b^2}{\sqrt{\gamma_\oplus - \hat{V}_\oplus^2 b^2}} \right]. \qquad (19)$$

where $b$ is the radius of the B's orbit around celestial object, and $\hat{V}_\oplus$ is the orbital angular velocity of the Earth around the celestial object (the Sun).

From (12), (13), (18), (19), and (4) we have,

$$\frac{\lambda_2^0 - \lambda_1^0}{\lambda_1} = h \left( \frac{\gamma_\oplus - \hat{V}_\oplus^2 b^2}{\gamma_* - \hat{V}_*^2 a^2} \right) \left( \frac{\hat{V}_\oplus - \hat{V}_*}{\alpha - h\,\hat{V}_\oplus} \right). \qquad (20)$$

This relation gives the difference in the red-shift between two points on the equator of the celestial object as observed from the Earth's surface. This general formula of the change of the red-shift for two vibrating atoms due to their positions depends on the parameter $h$. This parameter is related to the angle between the two lines of sight of the two atoms.

## 4. The Angle Between Two Null-Geodesics

In 4-dimensions, the angle between two null-vectors has no definition. We usually deal with photons in astronomical observations. The trajectory of such particles is a null-geodesic (whose tangent is a null-vector). Consequently, it will be impossible to try to measure the angle between any two null-geodesics. In fact, astronomers do measure the angle between the projections of such null-geodesics in a 3-dimensional hypersurface.

Let C be a point (an observer) in a 4-dimensional Riemannian space, at which the metric can be put in the form,

$$\begin{aligned} dS^2 &= g_{00}\, dt^2 - g_{ij}\, dx^i\, dx^j,\ (i, j = 1, 2, 3), \\ &= g_{00}\, dt^2 + a_{ij}\, dx^i\, dx^j. \end{aligned} \qquad (21)$$

Let $\Gamma_1, \Gamma_2$ be any two null-geodesics in the space (21) intersecting at the point C. If $\eta^\mu$ is the transport null-vector along the null-geodesic $\Gamma_1$, then we have,

$$g_{\mu\nu}\, \eta^\mu \eta^\nu = 0, \qquad (22)$$
$$g_{00}(\eta^0)^2 - g_{ij}\, \eta^i \eta^j = 0,$$
$$g_{ij} \frac{\eta^i \eta^j}{g_{00}(\eta^0)^2} = 1,$$
$$g_{ij}\, v^i\, v^j = 1. \qquad (23)$$

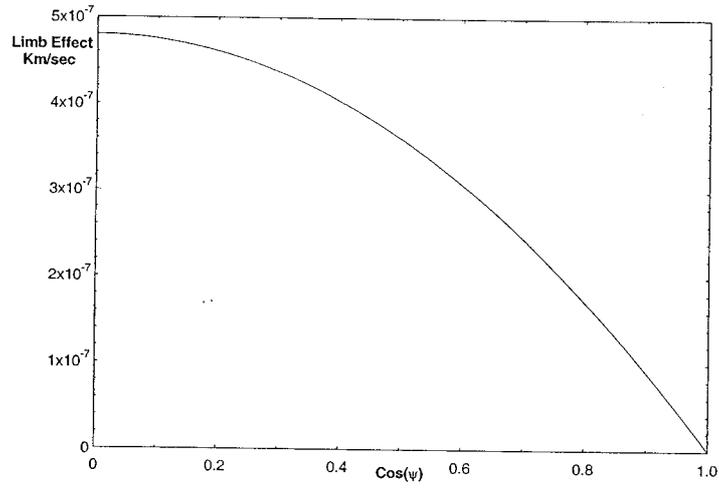

*Figure 2.* Theoretical limb effect

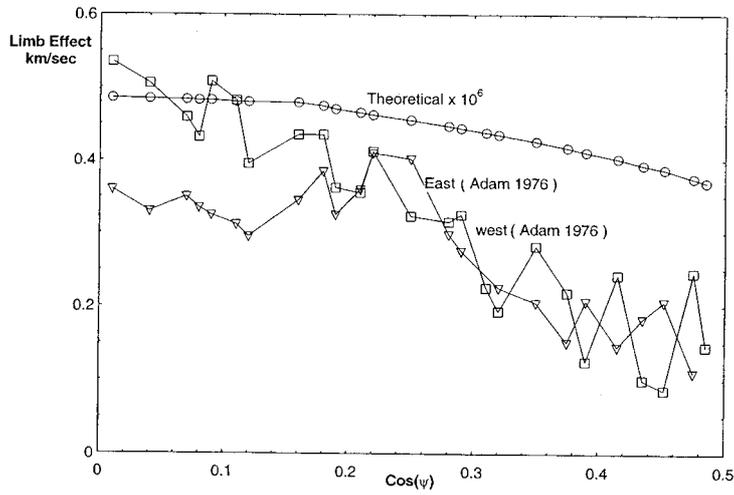

*Figure 3.* Observational and theoretical limb effect

where $v^i = \frac{\eta^i}{\sqrt{g_{00}}\,\eta^0}$, represents a unit vector in the 3-dimensional sub-space,

$$dl^2 = a_{ij}\, dx^i\, dx^j, \quad (i, j = 1, 2, 3). \tag{24}$$

If $w^i$ is the unit vector obtained in a similar manner, and representing the direction of the null-geodesic $\Gamma_2$ at C, then the angle between the two null-geodesics can be given as usual by:

$$\cos\theta = a_{ij}\, v^i\, w^j. \tag{25}$$

## 5. Gravitational Red-Shift Along The Sun's Disc

The formula obtained (20), in the present work, for the observed red-shift in the spectral lines of a celestial body is more general than previous treatments as it makes complete and direct allowances for the following essential factors:
 (i) the motion of the vibrating atom under consideration,
 (ii) the motion of the observer relative to the atoms,
 (iii) the direction of the line of sight used for the observations.

Besides, a direct estimation of the limb-effect is being given by considering the change in the observed wave-length purely due to the change in position of vibrating atom on the celestial body. Two identical atoms located at two different points on the equator of the celestial body, are observed by the same observer on the surface of the Earth. The difference between the two observed wavelengths has been calculated. One of the two atoms is considered to lie along the radial null-geodesic, joining the observer to the center of the celestial body, and the other one is observed obliquely. A direct relation between $\Delta\lambda/\lambda$ and the angle between the two lines of sight has been obtained.

By using the null-vectors defined by (13), (16) in (25), we find that,

$$h = \frac{\alpha\, b}{\sqrt{\gamma_\oplus}} \sin\theta, \tag{26}$$

Substituting in (20) we get the difference between the red-shifts of the two vibrating atoms at two different positions, on the equator of the Sun, in the form,

$$\frac{\lambda_0^2 - \lambda_0^1}{\lambda_1} = \frac{A_1 \sin\theta}{1 - B_1 \sin\theta},$$

where

$$A_1 = \frac{b}{\sqrt{\gamma_\oplus}}(\hat{V}_\oplus - \hat{V}_*)\left(\frac{\gamma_\oplus - \hat{V}_\oplus^2\, b^2}{\gamma_* - \hat{V}_*^2\, a^2}\right), \tag{27}$$

$$B_1 = \frac{b}{\sqrt{\gamma_\oplus}}\, \hat{V}_\oplus.$$

By using the transformation $\sin\theta = \frac{a}{b}\sin\psi$, then $A = \frac{a}{b} A_1$, $B = \frac{a}{b} B_1$, and the relation (27) takes the form,

$$\Delta Z = \frac{A\, \sin\psi}{1 - B\, \sin\psi}, \tag{28}$$

where $\psi$ is the angle between the projection of the radial null-geodesic $\Gamma_1$ in the spatial 3-dimensional space and the radius of the celestial object to the point $C_2$. It is clear that the formula (28) is free from the linear terms of the Einstein shift, so we expect that it gives rise to Doppler effect in the presence of the gravitational field and other effects which includes a type of interaction between velocity and the gravitational field. To eliminate the Doppler shift from the formula (28), we consider two atoms on the equator of the Sun symmetrically situated with respect to the angles $\psi, -\psi$, as shown in Figure 1a, and then we take the average value, we get,

$$\Lambda = \frac{A\ B\ \sin^2 \psi}{1 - B^2\ \sin^2 \psi}. \tag{29}$$

The relation (29) shows that $\Lambda$ takes the values,

$\Lambda = 0$ at the center of the Solar disk,

$\Lambda = \dfrac{A\ B}{1 - B^2}$ at the limb of the Solar disk,

and it increasing from center to the limb on both sides. So this formula may throw some light on the phenomenon of 'Limb Effect', observed in the Solar spectrum.

## 6. Concluding Remarks

The stellar limb effect was studied by many authors (e.g. Ayres et al., 1983; Ayers et al., 1988; Engvold et al., 1988). Also this phenomenon was studied by several authors in Solar spectrum by (St. John, 1928; Evershed, 1931; Adam, 1948, 1958, 1959, 1976, and 1979; Feldman et al., 1982; James and Keith, 1985; James et al., 1991; Hassler et al., 1991; Keith, 1992; Brekke, 1993; Achour et al., 1995; Brynildsen et al., 1995; Warren et al., 1998; Peter, 1999, and Peter and Judge, 1999). All those investigators detected the center to limb variation in the Solar spectrum along the Solar disk, except Hassler et al. (1991). Some of them tried to explain the physical processes which cause such phenomenon, but non of them was sure about the reason. Now in order to compare our theoretical result with observations we draw the curve showing the relation between $\Lambda$ and the disk position, for the Sun. It is found to be a parabola as shown in Figure 2. The general features of this curve are similar to those of the curve obtained from observation but with different scale as it is clear from Figure 3. The maximum value calculated from the relation (29) at the limb of the Sun ($\psi \simeq 90°$) is $\Lambda = 4.8 \times 10^{-7}$ kms sec$^{-1}$, while the corresponding average value obtained from observations is $\Lambda = 0.3$ kms sec$^{-1}$. This means that the observed values involves some other effects causing a magnification of the red-shift.

It is to be considered that the ratio between the theoretical and the observed values of $\Lambda$ is given by

$$\frac{\Lambda_{theo.}}{\Lambda_{obs.}} = 16 \times 10^{-7}. \tag{30}$$

Recently, some authors mentioned that the trajectory of a massless spinning particle (e.g. photon), in a background gravitational field, is a spin dependent (cf. Halpern, 1988; Wanas, 1998). The deviation from the null-geodesic motion is suggested to be due to a type of interaction between the quantum spin of the moving particle and the background gravitational field. There are some evidences for the existence of this interaction on the laboratory scale (Wanas et al., 1998, and 2000a), and on the galactic scale (Wanas et al., 2000b).

The coupling constant for this interaction is suggested to be the fine structure constant $\hat{\alpha} = \frac{1}{137}$. If we evaluate (30) using this constant, we find that,

$$\frac{\Lambda_{theo.}}{\Lambda_{obs.}} = \left(\frac{\hat{\alpha}}{6}\right)^2.$$

The question now is: Is there any impact of this interaction on the limb effect?.

Apart from the result obtained, we have got a byproduct which can be clarified in the following remarks:

1 – Although the angle between two null-geodesics is not defined theoretically, but it seems to be measured by observations e.g. the angle between the null-geodesics coming from center and limb of the Solar disk, i.e the semi-diameter of the Solar disc.

2 – Consequently, the angle between these two null-geodesics given by the relation (25), could be considered as a type of projection of the 4-dimensional null-geodesics on a 3-dimensional subspace.

3 – From (1), (2) one may conclude that:
'the angle between any two null-geodesics in 4-dimensional space cannot be defined theoretically, unless we project them in 3-dimensional subspace' and one may generalize this result in the following statement 'In the context of geometric field theories, calculations are carried out in 4-dimensional space, while observations (or experiment) are carried out in a $(3 + 1)$ dimensions.' This result has strong impact on the solution and interpretation of the geometric field theories including general relativity (Wanas, 1990).

## Acknowledgements

A.B. Morcos would like to thank Prof. J.E. Dyson, The Head of The Department of Physics and Astronomy, University of Leeds, U.K., for giving him all rights of using all the department facilities during his visit to the department.